\documentclass[%
 reprint,
 showkeys,
 amsmath,amssymb,
 prx,
floatfix,
]{revtex4-2}

\usepackage[utf8]{inputenc}
\usepackage[T1]{fontenc}
\usepackage{graphicx}
\usepackage{amssymb}

\usepackage[breaklinks]{hyperref}

\usepackage[per-mode=symbol]{siunitx}

\usepackage[noabbrev]{cleveref} 

\crefname{equation}{}{}
\crefname{figure}{Fig.}{Fig.}
\Crefname{figure}{Figure}{Figures}
\newcommand*{\refcite}[1]{ref.~\onlinecite{#1}}

\DeclareSIUnit{\cps}{cps} 
\DeclareSIUnit{\bit}{bit} 
\binoppenalty=\maxdimen
\relpenalty=\maxdimen

\begin{document}

\title[A scalable network for simultaneous pairwise quantum key distribution]{A scalable network for simultaneous pairwise quantum key distribution via entanglement-based time-bin coding}

\author{Erik Fitzke}
\affiliation{Institute for Applied Physics, Technische Universität Darmstadt,\\Schlossgartenstra\ss e 7,  64289 Darmstadt, Germany}

\author{Lucas Bialowons}
\affiliation{Institute for Applied Physics, Technische Universität Darmstadt,\\Schlossgartenstra\ss e 7,  64289 Darmstadt, Germany}

\author{Till Dolejsky}
\affiliation{Institute for Applied Physics, Technische Universität Darmstadt,\\Schlossgartenstra\ss e 7,  64289 Darmstadt, Germany}

\author{Maximilian Tippmann}
\affiliation{Institute for Applied Physics, Technische Universität Darmstadt,\\Schlossgartenstra\ss e 7,  64289 Darmstadt, Germany}

\author{Oleg Nikiforov}
\affiliation{Institute for Applied Physics, Technische Universität Darmstadt,\\Schlossgartenstra\ss e 7,  64289 Darmstadt, Germany}

\author{Felix Wissel}
\affiliation{%
Deutsche Telekom Technik GmbH, Heinrich-Hertz-Straße 3-7, 64295 Darmstadt, Germany
}%
\author{Matthias Gunkel}
\affiliation{%
Deutsche Telekom Technik GmbH, Heinrich-Hertz-Straße 3-7, 64295 Darmstadt, Germany
}%
\author{Thomas Walther}
\email{thomas.walther@physik.tu-darmstadt.de}
 \affiliation{Institute for Applied Physics, Technische Universität Darmstadt,\\Schlossgartenstra\ss e 7,  64289 Darmstadt, Germany}

\date{\today}

\begin{abstract}

We present a scalable star-shaped quantum key distribution~(QKD) optical fiber network. We use wavelength-division demultiplexing (WDM) of broadband photon pairs to establish key exchange between multiple pairs of participants simultaneously. Our QKD system is the first entanglement-based network of four participants using BBM92 time-bin coding and the first network achieving timing synchronization solely by clock recovery based on the photon arrival times. We demonstrate simultaneous bipartite key exchange between any possible combination of participants 
and show that the quantum bit error rate~(QBER) itself can be used to stabilize the phase in the interferometers by small temperature adjustments.
The key distribution is insensitive to polarization fluctuations in the network, enabling key distribution using deployed fibers even under challenging environmental conditions. We show that our network can be readily extended to 34~participants by using a standard arrayed-waveguide grating for WDM with~\SI{100}{\giga\hertz} channel spacing and that reconfigurable network connections are possible with a wavelength-selective switch. 
In a field test we demonstrate secure key rates of~\SI[separate-uncertainty = true]{6.3}{\bit\per\second} with a QBER of~\SI{4.5}{\percent} over a total fiber length of~\SI{108}{\kilo\metre} with~\SI{26.8}{\kilo\meter} of deployed fiber between two participants with high stability.

Our system features a relatively simple design of the receiver modules and enables scaling QKD networks without a trusted nodes to distances up to more than~\SI{100}{\kilo\meter} and to more than 100~users.
With such a network, a secure communication infrastructure on a metropolitan scale can be established. 

\end{abstract}

\maketitle

\section{Introduction}
The advent of quantum computers poses a risk for classical public-key cryptography~\cite{Grimes2019, Cheung_2008, Gerjuoy_2005}. 
One possible solution to this problem is quantum key distribution which uses quantum signals to share cryptographic keys between users~\cite{Gisin_2002, Scarani_2009, Xu_2020}. 
To date, a variety of QKD protocols, setups and testing links have been implemented and the achievable key rates and distances have continuously been increased~\cite{Xu_2020}. One major research direction has been to demonstrate long-distance QKD. Key exchange over hundreds of kilometers of optical fiber~\cite{Wang_2022_Twin_Field, Pittaluga2021, Chen_2020, ChenJP2021, Boaron_2018} as well as with satellite-based links spanning thousands of kilometers~\cite{Yin_2017, Liao_2018} have been demonstrated. Another focus of current research is the implementation of multi-user QKD networks often based on trusted nodes, i.e. relay stations set up between the network users that have full knowledge of the keys.

Currently, the largest of such networks is the Chinese QKD network connecting Beijing and Shanghai via a 2000~km long quantum backbone link that includes multiple metropolitan-area QKD networks~\cite{ChenYA2021}.
The big drawback of this trusted-node approach is that it is not applicable in situations where the users do not trust the network provider operating the central node. Alternative approaches to set up QKD networks not requiring trust in the central node can be realized by measurement-device-independent protocols or with entanglement-based protocols~\cite{Tang_2016, Joshi_2020, Alshowkan_2021, Xu_2020}. 

We have set up a star-shaped network for simultaneous pairwise QKD between multiple participants using time-bin entangled photon pairs. From a practical point of view, our system has some advantages over other approaches to QKD networks. For example, in order to  realize larger networks with prepare-and-measure two-party links based on weak coherent pulses and WDM, a wavelength-tunable sender module would be required for each participant. Our setup only requires a receiver module for each participant, which is beneficial especially when higher numbers of users are connected because it reduces the hardware complexity and cost.
A star-shaped network can also be realized with Measurement-Device-Independent~(MDI) or Twin-Field~(TF) QKD requiring additional fiber links for phase stabilization or locking of remote lasers~\cite{Liu_2019_Experimental, Minder_2019, Chen_2020, Pittaluga2021}. In multi-user networks the distances between users typically vary, essentially requiring the stabilisation of large unbalanced Mach-Zehnder Interferometers for such TF networks. In \refcite{Zhong_2022}, a ring-shaped multi-user TF network was realized to overcome these limitations. However, its scalability is limited in the number of participants and their distances since it requires the signals to be sent over one single fiber link passing through all participant's locations.

We employ dense wavelength division multiplexing~(DWDM) to realize QKD between multiple pairs of participants. Setups using DWDM for the distribution of polarization-entangled photons via optical fibers have been realized for photons generated by spontaneous parametric down-conversion~(SPDC) in periodically poled fibers~\cite{Zhu_2019} or crystals~\cite{Grice_2011, Kaiser_2014, Wengerowsky2018}.
Distribution of polarization-entangled photon pairs has been successfully implemented  over submarine fibers~\cite{Wengerowsky2019, Wengerowsky2020} and was the basis for a demonstration of a city-scale QKD network with eight users based on polarization coding~\cite{Joshi_2020}. QKD protocols using photon polarization require active polarization control because the birefringence in single-mode fibers can change over time, leading to large fluctuations of the initial polarization state. For fiber deployed underground, the polarization change in short fibers can be on a time scale of hours to days~\cite{Shi_2020}. However, substantially faster polarization changes have been observed in urban areas~\cite{Yoshino_13}. In \refcite{Ding_2017}, polarization fluctuations and their impact on QKD systems were systematically characterized and the required polarization tracking speed was measured to be on the order of multiple~\si{\radian\per\second} for inter-city and aerial links. Estimates show that polarization adjustments on a millisecond timescale are necessary for stable operation of a 68~km long aerial fiber QKD link~\cite{Liu_2018_aerial}.
Compensation schemes for the stabilization of the polarization drift have been proposed~\cite{Chen_2007,Xavier_2011,Li_2018}, but considerably increase the complexity of  QKD setups. Schemes re-adjusting the polarization based on the quantum bit error rate~(QBER) can only compensate relatively slow polarization changes. Depending on the key rate, some time is required in order to accumulate sufficiently many bits so that the QBER can be reliably estimated. Hence, QBER-based polarization stabilization may become infeasible for long transmission distances with low key rates and fast polarization changes.

In contrast to polarization-based QKD, protocols using the phase and arrival time of photons to encode qubits are very robust and independent of polarization changes. However, for phase-coding protocols, the critical parts of the setup that need stabilization are the interferometers,  which are set up in controlled environments at the photon source and the receiver stations, respectively.

Achieving stable operation is therefore independent of environmental influences on the transmission link. Thus, stable key exchange is even possible under relatively harsh environmental conditions such as urban areas where vibrations impair the polarization stability of the transmission link.

Therefore, we implemented a four-party quantum network using a time-bin protocol described in the next section. We achieve synchronization by clock recovery so that no separate synchronization channel is required. Our setup aligns the phases automatically and solely based on the QBER itself by tuning the interferometer temperatures. Employing a broad SPDC spectrum of a single photon pair source and DWDM in the optical C-band we demonstrate simultaneous key exchange between the four participants. Scaling the network to higher numbers of participants only requires connecting more receiver modules to the source. The simple design and building method of our receiver modules aids the scalability in the number of participants. It is readily scalable to 34~network users and compatible with DWDM multiplexing schemes that were previously used to establish QKD networks~\cite{Li_2016, Wengerowsky2018, Joshi_2020, Liu_2022}. With only slight modifications, it can even scale up to 102~participants.

\section{QKD protocol}
\label{sec:Protocol}
\begin{figure*}[tbp]
	\centering\includegraphics[width=15.1cm]{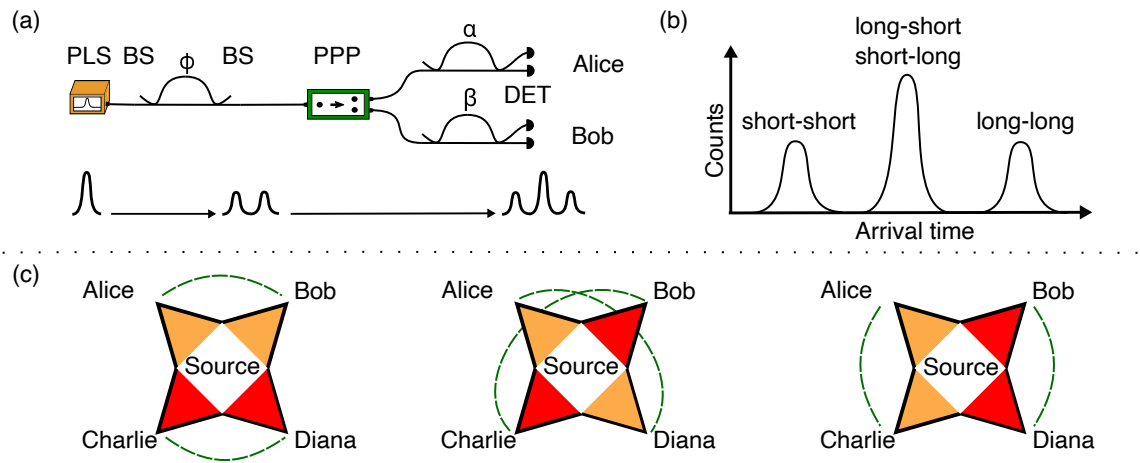}
	\centering\caption{(a)~Scheme for time-bin entanglement quantum key distribution with the BBM92 protocol~\cite{BBM92, Brendel_1999, Tittel_2000}. PLS:~Pulsed laser source, BS:~50:50 beam splitter, PPP:~Photon pair production, DET:~Single-photon detector. The symbols~$\phi$,~$\alpha$ and~$\beta$ indicate the phase delay of the interferometers. 
		(b)~Arrival time histogram at one of the detectors. Depending on the combination of long and short paths taken in the pump and receiver interferometers, the photons arrive in one of three time bins. Detection events in the early (short-short) and late (long-long) peak yield detections in the time basis while events in the central peak (long-short and short-long) yield detections in the phase basis.
		(c)~Possible configurations in a 4-participant star-shaped network with pairwise mutually exclusive combinations. For one given configuration, each participant is linked to exactly one other participant, so that two connected pairs can exchange keys simultaneously and independently. The dotted green lines indicate the  communication  over a classical channel, which is required in addition to the unidirectional quantum channel.}
	\label{fig:1}
\end{figure*}

We implement a time-bin variant of the BBM92 protocol~\cite{BBM92, Brendel_1999}. In the original implementation of the protocol, a photon pair source is placed between two participants, Alice and Bob~\cite{Tittel_2000}. Each participant holds a receiver module consisting of an imbalanced interferometer with two single-photon detectors, one at either output~(cf.~\cref{fig:1}(a)).

The basic idea of the protocol is as follows: In the photon pair source, pump pulses are sent through an imbalanced interferometer. The pulse duration is chosen to be shorter than the time delay, so that each pump pulse is split into a pair of non-interfering pulses. These pulses are then used to pump a nonlinear process such as SPDC or spontaneous four-wave mixing to produce entangled photon pairs. The pulse energy is chosen such that the mean number of photon pairs per pulse~$\mu \ll 1$, and therefore also the probability that a pulse generates more than one photon pair, is low.

Alice and Bob each receive one of the photons and detect it in one of three different time bins, as shown in \cref{fig:1}(b).
If the first laser pulse produces a photon pair, Alice and Bob detect their photons in the early or central time bin. Photons generated by the late laser pulse are detected in the central or late time bin. 
For detections in the early or late time bin Alice and Bob note down a 0 or 1, respectively. Detections in the central time bin are noted down as~0 or~1 depending on which of the two detectors registered the event. When the time delays of all three interferometers are matched, two-photon Franson interference~\cite{Franson_1989} leads to detection in correlated outputs for photons arriving in the central time bin.

In an ideal setup with perfectly indistinguishable interferometers, the probability for detection at two correlated outputs is given by~\cite{Tittel_2000, Marcikic_2004}
\begin{equation}
	P_{A_i,\,B_j}(\alpha,\beta,\phi) = \frac{1}{4}\left(1+(-1)^{i+j}\cos(\alpha+\beta-\phi)\right)\,
\end{equation}
with detector labels~$i, j\in {0,1}$. When the phases in the interferometers of Alice~($\alpha$), Bob~($\beta$) and the source~($\phi$) are aligned to~$\alpha+\beta-\phi = 2\pi n$ with~$n\in \mathbb{Z}$, Alice and Bob will always obtain the same bit values.
In this protocol, the two orthogonal bases required for QKD are the time basis consisting of measurements in the early and late time bin and the phase basis consisting of measurements  of the detector number in the central time bin.
In the key sifting step, Alice and Bob announce in which basis they measured the photon. If both detected a photon in the same basis, they append the corresponding bit value to the key. All events measured in different bases are discarded. 

Distribution of such time-bin entangled photons was realized over~\SI{50}{\kilo\metre}~\cite{Marcikic_2004} and~\SI{300}{\kilo\metre}~\cite{Inagaki_2013} and QKD using this scheme was implemented in a field test between two participants over~\SI{100}{\kilo\metre}~\cite{Honjo_2008}. Distributing time-~and wavelength-entangled photons with DWDM was previously demonstrated using photon sources based on spontaneous four-wave mixing in optical fibers~\cite{Li_2016} and silicon waveguides~\cite{Fang_2018}.

We extend wavelength-multiplexed entanglement distribution to a multi-user QKD network.

\section{Setup}

\begin{figure}[tbp]
	\centering\includegraphics[width=\columnwidth]{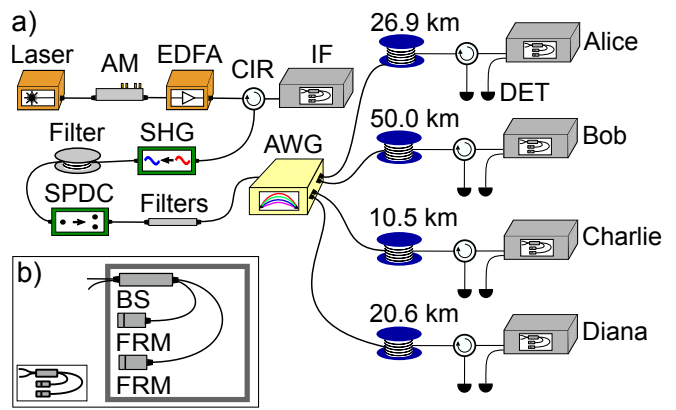}
	\caption{(a)~Setup for time-bin entanglement quantum key distribution with four participants. AM:~Amplitude modulator, EDFA:~Erbium-doped fiber amplifier, CIR:~Circulator, IF:~Interferometer, SHG:~Second harmonic generation, SPDC:~Spontaneous parametric down-conversion, AWG:~Arrayed-waveguide grating, DET:~Single-photon detector  (b)~Setup of the interferometers. BS:~50:50 beam splitter, FRM:~Faraday rotator mirror. All interferometers are placed in temperature-stabilized boxes.}
	\label{fig:2}
\end{figure}

Wavelength demultiplexing can distribute photon pairs by splitting the spectrum into different wavelength bins and sending them to more than two participants extending the key distribution scheme to a star-shaped multi-user network with the photon pair source at the center~(\cref{fig:1}(c)). In our experimental setup, depicted in \cref{fig:2}(a), we connect an arrayed-waveguide grating~(AWG) for wavelength-division demultiplexing to a photon pair source to realize this network structure.

In contrast to the idealized setup shown in \cref{fig:1}(a), we use Michelson interferometers with Faraday mirrors instead of Mach-Zehnder interferometers in order to remove the polarization dependence from our setup~(cf.~\cref{fig:2}(b))~\cite{Kersey_1991, Secondi_2004, Martinelli_1989}. Therefore, our setup does not require polarization stabilization.
We set up four identical receiver modules for the four participants Alice, Bob, Charlie and Diana to demonstrate simultaneous pairwise key exchange. The receiver modules are connected to the AWG via fiber spools of single-mode fiber with an attenuation of around~\SI{0.22}{\decibel\per\kilo\meter} typical for field deployed optical fibers.

The all-fiber design makes our setup compact and robust.
As pointed out earlier, phase stability of the interferometers is critical. Our phase stabilization is based on the precise temperature control of the interferometer temperatures. Due to the high temperature stability of approximately~$\SI{0.5}{\milli\kelvin}$, the phase adjustment can be solely based on the estimated QBER. In the following, we will briefly discuss the experimental details.

\subsection{Photon Source}

\begin{figure}[tbp]
	\centering\includegraphics[width=\columnwidth]{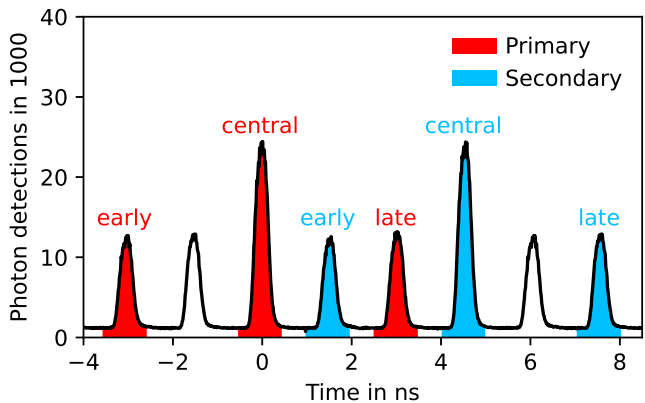}
	\caption{Histogram of photons traveling through~\SI{10.5}{\kilo\metre} of fiber before detection at one of Charlie's detectors. The histogram shows the pulse interleaving. The  structure of three peaks (early, central and late) for each color is produced by the interferometer time delay of~\SI{3.03}{\nano\second} (cf.~\cref{fig:1}~b). By setting the source repetition time to~\SI{4.55}{\nano\second}~(corresponding to~\SI{219.78}{\mega\hertz}), a secondary triplet of time bins (blue) is interleaved with the primary triplet (red). The data were accumulated over~\SI{90}{\second}. Although the peaks in the detection histogram are broader than the laser pulses of the photon source due to chromatic dispersion and detector timing jitter, they are sufficiently separated to avoid overlap, so that the QBER is kept low.}
	\label{fig:3}
\end{figure}

Photon pairs are created in a multi-stage process. The primary light source is a continuous-wave laser~(model Clarity, Wavelength~References) frequency locked to~\SI{1550.52}{\nano\metre}, i.e. between the channels~C33 and~C34 of the ITU~DWDM grid. A LiNbO\textsubscript{3}~amplitude modulator~(iXblue, \SI{10}{\giga\hertz}) shapes pulses with a repetition rate of~\SI{219.78}{\mega\hertz} and a width of about~\SI{300}{\pico\second}. The repetition time of~\SI{4.55}{\nano\second} is chosen such that there is no overlap between the time bins that belong to two successive pump pulses~(\cref{fig:3}). 

The pulses are amplified to an energy of up to~\SI{90}{\pico\joule} by an in-house-built erbium-doped fiber amplifier~(EDFA) and are passed through the source's Michelson interferometer. As our QKD protocol does not require the photon pair source to be trusted, no side channels are introduced by placing the interferometer before the SHG. A fiber-coupled PPLN crystal~(NTT Electronics) converts the wavelength to~\SI{775}{\nano\meter} via second harmonic generation~(SHG).
Tightly wound PM-780 fiber is used to filter out remaining light around~\SI{1550}{\nano\meter}. Two meters of such a coiled fiber result in a suppression of~\SI{79}{\decibel} and we use~\SI{5}{\meter} in our setup. A second PPLN crystal of the same type as above is used to generate energy-time entangled photon pairs in a \mbox{type-0 SPDC} process~\cite{Alibart2016, Aktas_2016}. High-pass filters remove the remaining \SI{775}{\nano\meter}-light.

In our setup, the source interferometer is placed before the SHG.
In the original implementation of the protocol, the source interferometer was set up directly before the SPDC crystal requiring a design for the second harmonic~\cite{Brendel_1999, Tittel_2000, Marcikic_2004}. In \refcite{Li_2016, Fang_2018}, SFWM instead of SPDC is used, which has the advantage that components at the telecom wavelengths can also be used for the pulse generation and pump interferometer. However, removing residual pump light is more complex due to the small frequency separation between the pump photons and the SFWM photons. Finally, in \refcite{Honjo_2008, Inagaki_2013}, rather than using an imbalanced interferometer, double pulses were produced at the telecom wavelength by employing two intensity modulators and were frequency-doubled for a SPDC process. With this scheme, the separation of the electronically generated double pulses must precisely match the interferometer delays and must not show timing jitter in order to obtain a low QBER.
By placing the interferometer before the SHG, we choose a different solution, but combine the advantages of these setups: only readily available components for the~\SI{1550}{\nano\meter}~wavelength range are required and the time delay can be precisely matched to the receiver interferometers as it can be built employing identical techniques (see \cref{ssec:receivers}). Nevertheless, the pulsed pump laser light can be efficiently separated from the second harmonic and conversely remaining~\SI{775}{\nano\meter} light can be efficiently removed from the photon pairs after the SPDC crystal.

The SPDC frequency spectrum is symmetric around the center frequency with a FWHM of approximately \SI{9.3}{\tera\hertz}~(\SI{75}{\nano\meter}). After passing a C-band filter transmitting all photons within a range of~$\pm\SI{2.55}{\tera\hertz}$ around the center frequency, the photons are distributed to the receivers by a standard telecommunication arrayed-waveguide grating~(AWG) with~\SI{100}{\giga\hertz} channel spacing. 

In order to set up a key exchange between a pair of participants, their fibers are connected to a pair of AWG channels that is symmetric around the center frequency and within the C-band filter's pass band. The number of available channels is limited by the lowest channel number of our AWG, which is C17. Therefore, 17~symmetric pairs of 100-GHz channels from C17~to~C50 are available for key exchange so that 34~participants can be connected to the network.

\subsection{Receiver interferometers}
\label{ssec:receivers}
In order to obtain low QBERs, the interferometers in the source and the receivers should be as similar as possible. Furthermore, the phase stability between the interferometers is of utmost importance. Therefore, we paid special attention to the design of the interferometers. 

The Michelson-type interferometers feature path length differences of~\SI{3.03}{\nano\second}. 
They consist of polarization-independent 50:50 beam splitters and Faraday rotator mirrors eliminating the sensitivity to birefringence in the interferometer arms~\cite{Kersey_1991, Secondi_2004, Martinelli_1989}
and enabling two-photon interference independent of the polarization state of incoming photons. The path combinations long-short and short-long in the pump and receiver interferometers need to be almost indistinguishable in order to achieve a sufficient two-photon interference visibility. Therefore, the differences between the time delays of the interferometers must be much smaller than the coherence lengths of the photon pairs.
We carefully manufactured our interferometers and monitored and corrected the path length deviations in each construction step. As we place the source interferometer before the SHG, the components of all five interferometers are identical. Thus, we were able to manufacture them all in a single attempt without resplicing. We then reduced the differences further to a few~\SI{10}{\micro\metre} by fiber expansion. These techniques enable the fast and precise manufacturing of larger numbers of interferometers required to achieve scalability of our approach.

Each interferometer is enclosed by a box to shield it from environmental temperature fluctuations. The box temperature can be adjusted with thermo-electric elements driven by digitally controlled temperature controllers developed in-house. The sensitivity of the interferometer phase to temperature changes is in the range of~\SIrange[parse-numbers=false]{0.033\pi}{0.045\pi}{\per\milli\kelvin} and the box temperature can be adjusted with a precision of~\SI{0.5}{\milli\kelvin}.

Chromatic dispersion in fibers broadens the peaks in the arrival time histogram.

If the time delay in the interferometers is chosen too small, chromatic dispersion will lead to overlapping peaks in the photon arrival time histograms and thus to an increased QBER for transmission over long fibers. 
For a two-user system, the repetition rate and time bin width can be optimised for a particular transmission distance, but for a multi-user system, the maximum distance from the source to a participant limits the maximum repetition rate. 
In principle, dispersion compensation modules can be used, but they increase the insertion loss. As our goal is to keep the receiver modules as simple as possible for the sake of scalability in the number of users, we opted for an implementation without dispersion compensators. Therefore, we chose a time delay of approximately~\SI{3}{\nano\second}. However, a large time delay limits the pulse repetition rate. 
A natural choice for the repetition time of the pulse generator would be three times the interferometer time delay, resulting in equidistant peaks in the arrival time histogram.
For the fiber lengths used in this paper, the broadening from chromatic dispersion is relatively small leaving relatively large unused gaps between the peaks in the arrival time histogram. In order to efficiently use these gaps, we interleaved consecutive repetition cycles by setting the repetition time of the pulse generator to 3/2~of the interferometer time delay~(cf.~\cref{fig:3}), thereby effectively lifting the constraint that large time delays would otherwise impose on the maximum repetition rate. To the best of our knowledge, this is the first demonstration of such an interleaving technique to increase the effective repetition rate for comparatively large interferometer time delays. 

\subsection{Data acquisition and phase calibration}
\label{ssec:data_acqusition}

We have extensively automated the setup's operation. Data acquisition is split into 90~second long runs with approximately 6~second long intermissions for qubit evaluation.
The photons are detected by avalanche single-photon detectors~(ID~Quantique~ID220) with a timing jitter of about~\SI{250}{\pico\second}. All detectors are set to~\SI{20}{\percent} quantum efficiency with a dead time of~\SI{10}{\micro\second}. Timestamps are recorded by time taggers~(ID~Quantique~ID900) with~\SI{13}{\pico\second}~resolution. For the experiments presented in \cref{sec:results} we synchronized the time taggers by sharing a~\SI{10}{\mega\hertz} clock signal and in \cref{sec:Towards_separated_parties} we show that synchronization can also be achieved by clock recovery from the photon arrival times.

A common time reference~$t_0=0$ needs to be established, so that the participants can assign their detected photons correctly to the events registered by their counterpart.
We establish~$t_0$ by determining the maximum of the cross-correlation of detection events in the first run.  
For a key exchange between distant users, this would require a public comparison of all photon arrival times of this run. Consequently, the data from the first run cannot be used to generate key bits. However, using the data from the first run to establish~$t_0$ has the advantage that there is no need for a separate alignment phase. Furthermore, for our offset alignment procedure, it is neither necessary to stop data acquisition for recalibration nor additional components are required.

Temperature changes of the link fibers can cause an arrival time drift due to a change of the optical path lengths. Our system can automatically compensate for drifts up to~\SI{2.2}{\nano\second} per run, i.e. per~\SI{90}{\second}. The thermal sensitivity of the propagation delay in single-mode fibers around~\SI{1550}{\nano\meter} has been reported to be around~\SI{39}{\pico\second\per\kilo\meter\per\kelvin}~\cite{Bousonville_2012, Hartog_1979, Slavik_2015}. For the maximum transmission distance of almost~\SI{77}{\kilo\meter} we consider in~\cref{sec:results}, the system could thus compensate heating rates of more than~\SI{0.7}{\kelvin} per~\SI{90}{\second} acting on the whole transmission link simultaneously. Typical temperature change rates affecting the transmission link should be much smaller on this time scale, especially for fibers deployed underground. Note that in~\cref{sec:Towards_separated_parties} we introduce clock recovery, which also lifts this limitation.

In addition to the arrival time calibration, the interferometer phases are automatically calibrated such that a minimal QBER is achieved. In the 6~second-long intermissions between runs, the QBER is estimated automatically every three minutes and the box temperatures are adjusted such that a minimal QBER is attained. 

This whole startup takes at most 45~minutes due to the heat capacity of the boxes and is completed significantly faster if the interferometers are already pre-aligned. The box design comes with a trade off: A high heat capacity limits the alignment speed, but stabilizes the interferometer against fast temperature fluctuations. With our interferometer design, we opted for a trade-off providing both sufficient temperature stability and an acceptable alignment speed.

After the startup, key exchange can commence. The QBER is estimated automatically every three minutes in order to detect interferometer phase drifts and the interferometer temperatures are adjusted automatically in order to keep the QBER as low as possible. In the current setup, the timestamps are processed on the same computer. 
When the users are spatially separated, a randomly sampled fraction of the sifted key is made public to estimate the QBER for the postprocessing steps~\cite{Gisin_2002, Xu_2020}. This QBER information can simultaneously be used to align the interferometers, so that no further security risks or information leakage results from adjusting the phases based on the QBER.

\section{Results}
\label{sec:results}
\begin{figure}[tbp]
	\centering\includegraphics[width=\columnwidth]{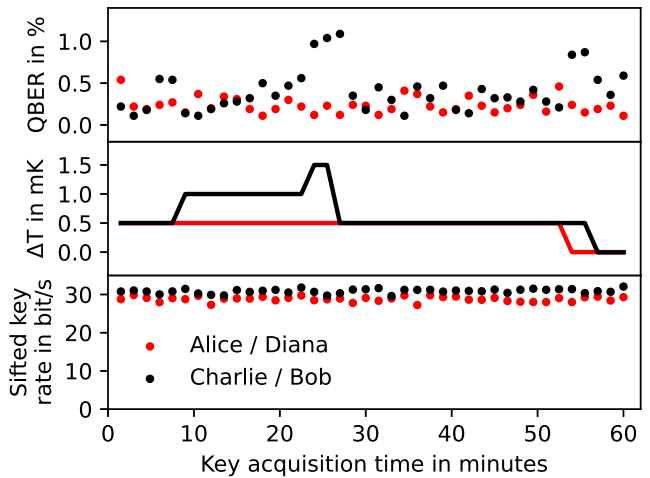}
	\caption{A simultaneous key exchange for Alice~/~Diana~(red) and Charlie~/~Bob~(black) without fiber spools and with a mean photon pair number per pulse~$\mu$ below~\num{1e-3} demonstrating low quantum bit error rates~(QBER). The temperature change~$\Delta $T shows adjustments made at Alice's and Bob's receivers to minimize the error rate. Each data point was obtained from one run of~\SI{90}{\second}. The average QBER was~\SI{0.24}{\percent} for Alice~/~Diana and~\SI{0.41}{\percent} for Charlie~/~Bob, respectively.}
	\label{fig:4}
\end{figure}

The QBER during key exchange with our setup strongly depends on the mean number of photon pairs per pulse~$\mu$. For values of~$\mu$ below~\num{1e-3}, low QBERs are expected because the emission of multiple photon pairs per pulse becomes highly unlikely. In order to assess the achievable quality of the correlations in the phase basis, measurements with an average SHG power of~\SI{0.75}{\micro\watt} were performed
without additional fiber spools between the source and the receivers. \Cref{fig:4} shows a one-hour long key exchange for such low values of~$\mu$. An average QBER of~\SIlist{0.24;0.41}{\percent} was reached for the combinations Alice~/~Diana and Charlie~/~Bob.
The QBER between Charlie and Bob shows maxima around~28~minutes and 55~minutes which were caused by phase drifts in the respective interferometers during the measurement. However, the automatic phase calibration compensated for the drift and the QBER quickly returned to lower values. Since the QBER is a symmetric error function yielding no information regarding the direction of the phase drift, the algorithm occasionally adjusts the temperature in the wrong direction, as seen for the peak at 28 minutes. However, the algorithm quickly recognized the wrong decision and corrected it automatically. Between 30~minutes and 50~minutes, no temperature adjustments were necessary for either party. Key exchange between Alice/Diana did not require any temperature adjustments for more than 50~minutes.
The very low error rates show that the losses and time delays in the arms of our interferometers are well matched. 

The reliability of our phase stabilization algorithm was estimated in a long-term measurement over several hours of continuous operation. For these measurements, the average number of photon pairs per pulse~$\mu$ created in the frequency range of one channel pair was increased to be in the range of~0.03 corresponding to an average SHG pump power of~\SI{30}{\micro\watt} to maximise the average secure key rate. For this mean photon number, the probability for a pulse to produce a photon pair is greatly increased. For even higher values of~$\mu$ the probability to detect photons from different pairs due to multi-photon pair emission becomes relevant for the QBER. 

\begin{figure}[tbp]
	\centering\includegraphics[width=\columnwidth]{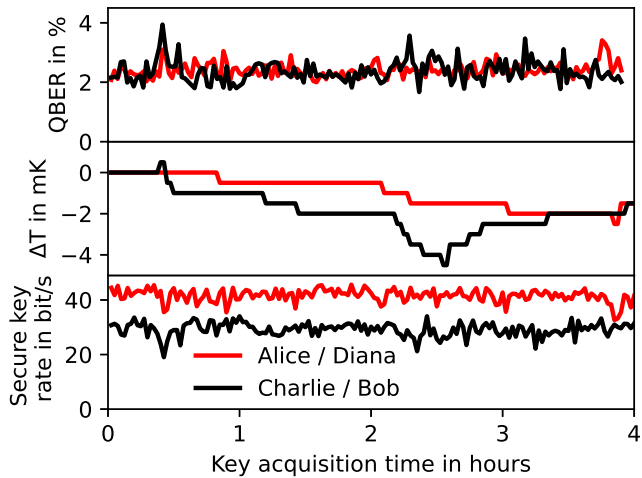}
	\caption{Estimated secure key rates and QBER for simultaneous key exchange over fiber spools with a total length of~\SI{47.5}{\kilo\meter} for Alice~/~Diana~(red) and~\SI{60.5}{\kilo\meter} for Charlie~/~Bob~(black) with a $\mu$ in the range of~0.03. $\Delta$T~denotes the temperature adjustments made at Alice's and Bob's receiver interferometers. Each data point represents one 90~second long run.}
	\label{fig:5}
\end{figure}

A four-hour long key exchange between the pairs Alice~/~Diana and Charlie~/~Bob with average QBERs of~\SI{2.41}{\percent} and~\SI{2.36}{\percent} is shown in \cref{fig:5}. As before, the automatic phase stabilization algorithm adjusts the interferometer temperatures such that the error rates stay at a minimum. Even though the temperature of the interferometers can be adjusted every three minutes, the error rates remained stable over much longer periods. As the QBER for each run is calculated over the duration of the run (\SI{90}{s}), phase fluctuations occurring during a run would increase its QBER. Similarly, slower phase drifts would lead to a trend in the QBER showing up for multiple runs. The QBER during the entire measurement stays low and neither interferometer temperature nor QBER show a significant drift over time, demonstrating the excellent short-~and long-term phase stability of our system. If the secure key rate dropped to zero, the key exchange would have to be paused until stable operations would become possible again. However, the secure key rates in \cref{fig:5} never drops to zero, i.e. keys were exchanged without interruption.

The sum of the fiber lengths from the source to Alice and from the source to Diana, i.e the effective key transmission distance, was~\SI{47.5}{\kilo\metre} and an average sifted key rate of~\SI{70}{\bit\per\second} was achieved. Between Charlie and Bob the effective distance was~\SI{60.5}{\kilo\metre}~(cf.~\cref{fig:1}(a)) with a sifted key rate of~\SI{49}{\bit\per\second}.
Assuming error reconciliation based on low-density parity-check codes, we estimated the secure key rates~$r_\textrm{sec}$ which can be calculated from the sifted key rate~$r_\textrm{sift}$, the QBER~$q$ and the reconciliation efficiency $f$~(cf.~\refcite{Elkouss_2009}):
\begin{equation}
	r_\textrm{sec}= r_\textrm{sift}\Big(1-\big(1+f\big)\big(-q\log_2(q)-(1-q)\log_2(1-q)\big)\Big)\,.
\end{equation} 
Using a conservative estimate for the reconciliation efficiency of~$f=\num{1.5}$~(cf.~\refcite{Elkouss_2009}), we estimated the average secure key rate as~\SI[separate-uncertainty = true]{42\pm 3}{\bit\per\second} and~\SI[separate-uncertainty = true]{29\pm 3}{\bit\per\second} for Alice~/~Diana and Charlie~/~Bob.

\begin{figure}[tbp]
	\centering\includegraphics[width=\columnwidth]{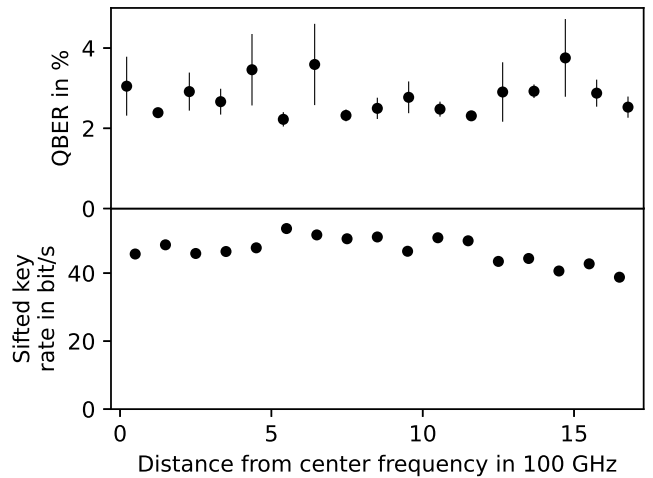}
	\caption{Quantum bit error rate~(QBER) and sifted key rate for key exchange between Charlie and Bob over~\SI{60.5}{\kilo\metre} of fiber for different AWG channel pairs symmetric around our center frequency. Each data point was averaged over 8~consecutive 90-second runs. The error bars represent the standard deviation of the QBER for each channel. The error bars of the sifted key rate are so small that they are not visible. }
	\label{fig:6}
\end{figure}

One main advantage of our system is the scalability of the photon source with respect to the number of users. In order to demonstrate that all~34 available channels are usable for key exchange, we connected the pair Charlie~/~Bob over a distance of~\SI{60.5}{\kilo\metre} of fiber and measured the QKD performance with each of the~17~available channel pairs (cf.~\cref{fig:6}). All channel pairs in the pass band of the C-band filter offer a similar performance. Thus, with a channel spacing of~\SI{100}{\giga\hertz} of the AWG, we conclude that 34~participants can be connected to our source simultaneously.

\begin{figure}[tbp]
	\centering\includegraphics[width=\columnwidth]{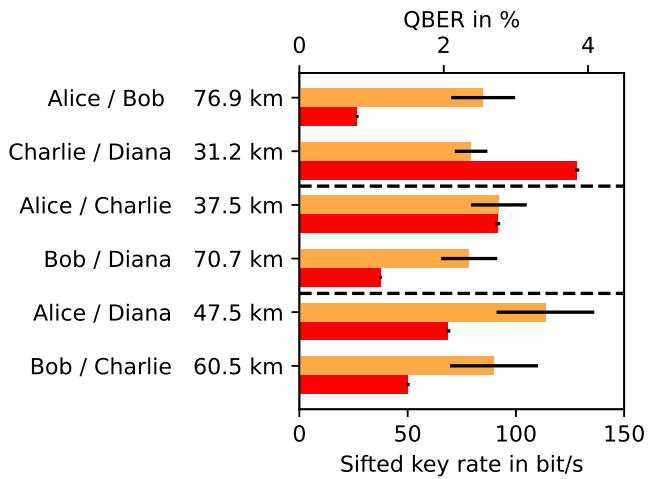}
	\caption{QBER~(orange) and sifted key rates~(red) for all combinations of the four participants Alice, Bob, Charlie and Diana 
		over different fiber lengths. Fiber spools were assigned as shown in~\cref{fig:2}(a). The data was obtained during three 20-run measurements via simultaneous pairwise key exchange. The dashed lines separate the results of the three measurements. The error bars represent the standard deviation of the results for each combination.}
	\label{fig:7}
\end{figure}
In order to assess the capabilities of our system over distances typical for metropolitan quantum networks, the performance of the network was investigated by connecting the AWG and the receivers with fibers of different lengths for each participant~(cf.~\cref{fig:2}(a)). In this case, the communication between different pairs of participants is set up by reconfiguring the AWG connections and recalibrating the receiver interferometer phases. This is done via the same startup procedure described above.
We tested all possible network configurations shown in \cref{fig:1}(c) and demonstrated successful key exchange between all combinations of all four participants, as shown in \cref{fig:7}. 
The results were obtained with the same mean photon pair number per pulse~$\mu$ for all combinations. Hence, a simultaneous key exchange over total distances between~\SI{31.2}{\kilo\metre} and~\SI{76.9}{\kilo\metre} was achieved. 

\section{Field-Test: Towards a Distributed QKD-Network}
\label{sec:Towards_separated_parties}
\begin{figure}[tbp]
	\centering
	\includegraphics[width=\columnwidth]{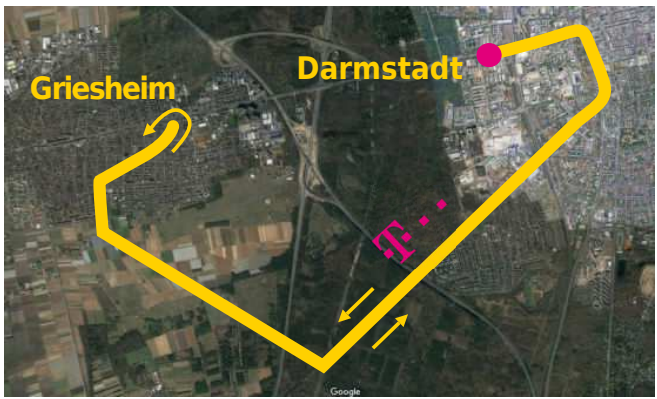}
	\caption{Deployed standard single mode dark fiber link of Deutsche Telekom for the field test. The link consists of a looped fiber running from the Deutsche Telekom facility in Darmstadt to Griesheim and a parallel fiber running back to Darmstadt. The fiber length is~\SI{26.8}{\kilo\meter} with a total loss of~\SI{6.7}{\decibel}. (Google Maps 2021 \copyright )}
	\label{fig:8}
\end{figure}

The results presented above demonstrate the feasibility and advantages of our approach to QKD networks. In order to address typical challenges to be overcome on the way towards a real-world multi-user distributed network, first results of a field test with our system are reported in this section.  The entire system was moved to a facility of Deutsche Telekom where Alice was connected via a~\SI{26.8}{\kilo\meter} deployed dark fiber loop running from the city of Darmstadt to Griesheim and back to Darmstadt (cf.~\cref{fig:8}). Bob, Charlie and Diana were connected via~\SIlist{81.2;9.6;20.9}{\kilo\meter} of spooled fiber. Compared to \cref{ssec:data_acqusition}, we changed the operation mode of the data acquisition: A measurement is still split into~\SI{90}{\second} long runs, but the evaluation of one run is performed while the next run is recorded without intermissions so that the transmission time is used most efficiently for qubit exchange.
We now address two further challenges on the way towards a distributed QKD network using our approach: channel reconfiguration and clock synchronization.

\subsection{Automatic channel reconfiguration}

In our setup described above, a communication with fixed channel allocations requires manual reconfiguration of the AWG when different pairs of participants want to exchange quantum keys. Wavelength-division multiplexing schemes as presented in \refcite{Wengerowsky2018, Joshi_2020} can be used to implement fully-connected networks with our setup, since they also use broadband photon sources. Moreover, in combination with active or passive time multiplexing, networks with participants grouped in fully connected sub-networks or fully connected networks can be realized~\cite{Liu_2020, Fang_2018}. Recently, a fully-connected network with 40 users using time-~and wavelength multiplexing was presented in~\refcite{Liu_2022}.

An alternative to such schemes with a fixed channel allocation are networks employing a wavelength-selective switch (WSS)~\cite{Lingaraju_21, Appas_2021, Alshowkan_2021} for demultiplexing of the photon pair spectrum. In contrast to an AWG, a WSS can be reconfigured electronically. It would allow to define which parts of the spectrum are routed to which users, enabling routing and allocation of bandwidth tailored to the participant's key demands and enabling dynamic adaption when the demands change over time. In order to show that our system can readily be used with a WSS, we replaced the AWG by a WSS for the field test.

\subsection{Synchronization by clock recovery} 

Another challenge is the synchronization of the clocks of the photon source and the receiver modules when they are located at distant locations. 
A variety of approaches exist to achieve synchronization between distant QKD modules.
For example, synchronization can be achieved via a dedicated synchronization channel. Such a channel can be implemented by employing a separate fiber or by wavelength-~or time-multiplexing of synchronization signals with the photons used for QKD in the same fiber~\cite{Bienfang_2004, Tanaka_2008, Williams_2021, Islam_2017_provably, Walenta_2014}. Another approach is to rely on stable local clocks and linking them to an external time reference such as GPS~\cite{Ursin2007, Scheidl_2009, Ecker_2021}. However, this makes the system prone to Denial-of-Service attacks by an attacker who has access to the reference signal.
All of these approaches have the disadvantage that they require additional resources such as a dedicated classical channel or hardware such as GPS clocks, complicate the setup or reduce the achievable key rate by using time slots that can therefore not be used for qubit exchange. 

An alternative is to perform clock recovery on the photon arrival times~\cite{Calderaro_2020, Agnesi_20, Cochran_2021, Wang_2021_synchronization}. 
Clock synchronization  between two distant stations receiving entangled photons has been demonstrated by evaluating the cross-correlation of detections~\cite{Valencia_2004, Ho_2009}. Clock recovery based on the arrival times of non-entangled photons was implemented and named Qubit4Sync in \refcite{Calderaro_2020} and applied in a QKD system in \refcite{Agnesi_20}. Frequency recovery from photon arrival times for satellites was investigated in \refcite{Wang_2021_synchronization}.

In our system, the time-bin based BBM92 protocol leads to an arrival time distribution of the photons with the periodicity of the photon source repetition time (cf.~\cref{fig:3}). By analyzing the photon arrival times, the clock speed of the source can be retrieved. This approach requires neither additional hardware nor sacrificing additional qubits for synchronization purposes.

In order to demonstrate QKD with clock recovery, we set up a separate time tagger for each participant. The time tagger of Diana provides a~\SI{10}{\mega \hertz} clock signal to the photon source so that the clock stability of the source is essentially that of Diana's time tagger. The time taggers of Alice, Bob and Charlie are not connected to any reference clock. For synchronization between the participants we performed clock recovery on the time stamps with a self-developed algorithm. The clock recovery for the participant is completely independent of each other and no further data exchange between the participants is required. Note that employing clock recovery also lifts the requirement from \cref{ssec:data_acqusition} that the propagation delay drift due to thermal expansion of the link fiber must be less than~\SI{2.2}{\nano\second} per run. Drifts of the propagation delay are compensated along with clock drifts.

The reliability of the clock recovery depends on the intrinsic stability of the clocks of the receiver and the photon source. Clearly, recovering the clock reliably becomes more challenging for less stable clocks and smaller photon arrival rates. Consequently, clock recovery is most challenging for Bob because his transmission link is the longest. Due to the high losses, Bob only measured a mean count rate of around~\SI{9700}{\cps}. For such low count rates, the clock recovery algorithm occasionally slips by one or multiple time bins and a sudden increase in the time basis QBER is the consequence. Such an event is automatically detected by our algorithm and the time reference is then recalibrated by cross-correlation analysis of the current run to re-establish synchronization as it was described in \cref{ssec:data_acqusition} for the initial determination of the time reference~$t_0$. A detailed discussion of our clock recovery algorithm is beyond the scope of this paper and will be presented elsewhere.

For our synchronization scheme it is neither necessary to stop data acquisition nor are additional components required. Thus, by combining clock recovery and the reference agreement on~$t_0$ by the initial cross-correlation, the system can be operated without an additional high-accuracy timing synchronization channel. 

\subsection{Results of the field test }

\begin{figure}[tbp]
	\centering
	\includegraphics[width=\columnwidth]{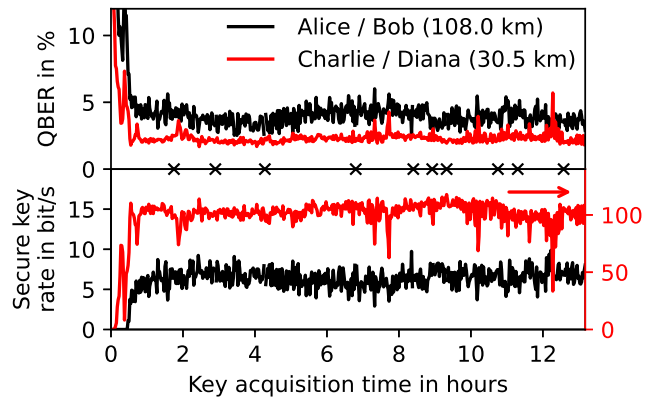}
	\caption{QBER and secure key rate acquired during the field test. A wavelength-selective switch is employed for demultiplexing of the SPDC spectrum. Alice is connected via the deployed fiber (\SI{26.8}{\kilo\meter}, see~\cref{fig:8}). The other participants are connected via spooled fiber: Bob via~\SI{81.2}{\kilo\meter}, Charlie via~\SI{9.6}{\kilo\meter} and Diana via~\SI{20.9}{\kilo\meter}. The total distance between two users is the sum of the individual link lengths. Diana's time tagger is synchronized to the photon source while for the other participants synchronization is achieved via clock recovery from the photon arrival times. Black crosses indicate runs where Bob's clock recovery failed and the delay to Alice was therefore automatically recalibrated via cross-correlation.}
	\label{fig:9}
\end{figure}

Measurement results of the field test including the modified setup using the WSS, clock recovery as well as the deployed fiber are shown in~\cref{fig:9}. The WSS was set up so that Alice and Bob each receive a bandwidth of~\SI{50}{\giga\hertz} of the photon pair spectrum while Charlie and Diana each receive~\SI{25}{\giga\hertz}. This choice represents a scenario where Alice and Bob require a higher key rate and therefore obtain a wider part of the SPDC bandwidth. Note that in order to obtain mean photon pair numbers per pulse and channel pair~$\mu$ comparable to the measurements with the \SI{100}{\giga \hertz} AWG, we increased the SPDC pump power from~\SI{30}{\micro\watt} to~\SI{90}{\micro\watt}.

It can be seen in~\cref{fig:9} that after the startup phase of approximately~\SI{45}{\minute} the sifted key rate and QBER are stable over hours. Disregarding the startup phase, the QBER between Alice and Bob is with~\SI[separate-uncertainty = true]{4.5\pm 2.0}{\percent} slightly higher than the QBER of~\SI[separate-uncertainty = true]{2.6\pm 1.2}{\percent} between Charlie and Diana. This is a consequence of the larger channel width for Alice and Bob resulting in a higher probability for multi-photon pair emission. 
A few spikes in the QBER between Charlie and Diana are visible in the second half of the measurement. They are caused by phase instabilities which were quickly compensated for by the automatic phase alignment algorithm.
The secure key rate and standard deviation after the startup phase are~\SI[separate-uncertainty = true]{6.3\pm 1.1}{\bit\per\second} between Alice and Bob and~\SI[separate-uncertainty = true]{102\pm 8}{\bit\per\second} between Charlie and Diana. Over the measurement time of more than~\SI{13}{\hour}, only 10 automatic recalibrations of the clock time by cross-correlation were performed between Alice and Bob and none for Charlie and Diana. This means that despite of the high losses in the link to Bob, the clock recovery algorithm slipped in less than~\SI{2}{\percent} of the runs, demonstrating its excellent stability.

\section{Discussion and Outlook}

Our QKD system proved to be able to distribute quantum keys between any two pairs of participants simultaneously without requiring a trusted node. The high stability of the QBER is reflected in the low standard deviation of the secure key rates. These results demonstrate the stability improvement of time-bin entangled protocols versus polarization based protocols in comparable networks~\cite{Joshi_2020}. We attribute this increased stability to the fact that our protocol is insensitive to environmental effects impairing the polarization stability of the transmission fiber. For the same reason, we expect that our system, when used in the field with deployed fiber, can achieve a stability comparable to the performance demonstrated with fiber spools. This will enable QKD even in challenging environmental conditions, such as transmission via aerial fiber or through densely populated urban areas. In a field test with Alice connected to more than~\SI{26}{\kilo\meter} of deployed fiber and Bob connected to more than~\SI{81}{\kilo\meter} of spooled fiber, we have verified that the QBER indeed exhibits excellent stability even for a total fiber length between the users of~\SI{108}{\kilo\meter}.

We demonstrated wavelength demultiplexing with an arrayed-waveguide grating (AWG) and with a wavelength-selective switch (WSS). The WSS will enable dynamical bandwidth allocation for the participants based on the key demands in the network. We showed that we can reduce the channel widths at least down to~\SI{25}{\giga\hertz}  with the WSS and that we can route different bandwidths to different pairs of participants. We showed that as an alternative to the WSS, an arrayed-waveguide grating with~\SI{100}{\giga\hertz} channel spacing can be used to connect 34~participants in 17~pairs. However this AWG only uses a span of~\SI{3.4}{\tera\hertz} the SPDC spectrum. 

If the full width of our C-band filter was used, QKD would be possible for ITU channels~8 to~59. Thus, up to 102~participants could be connected to our photon source for simultaneous pairwise key exchange with a suitable~\SI{50}{\giga\hertz} demultiplexer. AWGs with~\SI{50}{\giga\hertz} channel spacing are commercially available and even lower channel width and thus higher numbers of participants would be possible with a wavelength-selective switch, given it has sufficiently many output ports. 

We demonstrated timing synchronization solely based on the evaluation of the photon cross-correlation and local clock recovery at the receivers.  
In order to operate the network with remote participants (Alice and Bob, for example) all they have to do is to reset the clocks of their time taggers roughly at the same time. 
We evaluated the cross-correlation in a range of~$\pm\SI{2.5}{\milli \second}$ which was enough to cover all optical and electronic delays between participants in our setup. However, given sufficient computing power, the cross-correlation could be evaluated in the range of seconds. Alice and Bob would then only need to agree on the starting time with a precision of milliseconds to seconds which can be easily achieved for example via classical network communication and the NTP~protocol~\cite{Mills_1991}.
Alice then sends the time stamps of all her detection events in the first run to Bob, who calculates the cross-correlation with his detected events and deduces the time shift between his clock and Alice's. The time stamps of the first block are then discarded. All clock drifts are determined locally by Alice and Bob solely based on their own detection events. High-precision clocks or synchronization signals are not required with our scheme. Especially, the classical channel is not timing critical and could instead be realized e.g. via regular communication over the internet. 

Clock recovery is one of our measures to make the receiver modules simpler and  cheaper. Due to the chosen QKD protocol itself, the modules do not comprise active components such as phase modulators or polarization controllers. All in all, due to the relatively simple design, manufacturing of higher numbers of such modules for networks with dozens of users becomes feasible.

The key rates of our setup are currently limited by the repetition rate of our source and the detection efficiency and dead time of our avalanche photo diodes. 
For example, commercial superconducting nanowire single-photon detectors~(SNSPD) can reach polarization-independent efficiencies greater than~\SI{70}{\percent}
with full recovery times around~\SI{60}{\nano\second}~(e.g.~IDQ~ID281)~\cite{id281_brochure}. 
Using such detectors could increase the key rate by a factor of at least 12.3, solely by improving the detection efficiency from~\SI{20}{\percent}~to~\SI{70}{\percent}. 
In addition, the limit on the key rate imposed by our detector dead time of~\SI{10}{\micro\second} can be overcome with SNSPDs due to their short recovery time. An increase of the repetition rate for a constant mean photon  pair number per pulse~$\mu$ will lead to an almost proportional increase of key rates when the detector dead times are short compared to the mean arrival time between photons.

The width of the time bins could be reduced further by using shorter pump pulses, dispersion compensation and a detection setup with less jitter. Our pulse interleaving technique (cf.~\cref{fig:3}) could then be extended by setting the pulse generator's repetition time to $3/2^n$ times the interferometer delay without requiring any further components. 
It is reasonable to assume that a width of~\SI{95}{\pico\second} per time bin is sufficient to accommodate short source laser pulses and detector timing jitter when SNSPDs and fast high-precision acquisition electronics with a timing jitter as low as~\SI{25}{\pico\second}~\cite{id281_brochure} are used. Thus, the repetition rate could be increased by a factor of 16 to approximately~\SI{3.5}{\giga\hertz} by interleaving 32~pulses without overlap between time bins. 

The increase in efficiency and repetition rate will result in an overall improvement of the sifted key rate by a factor of approximately~197 compared to the data presented here, i.e. sifted key rates above~\SI{9}{\kilo\bit\per\second} over a distance of~\SI{60}{\kilo\metre} of standard telecommunication fiber are feasible. 
While our setup in its current configuration is already suitable for communication over metropolitan-area distances, SNSPDs and higher repetition rates can conversely be used to increase the transmission distances significantly beyond the~\SI{108}{\kilo\metre} we have demonstrated. 

\section{Conclusion}
We presented for the first time an all-fiber time-bin entanglement-based quantum key distribution system enabling simultaneous QKD between any two pairs of participants by employing wavelength division multiplexing.
Simultaneous key exchange was demonstrated over fiber lengths up to~\SI{108}{\kilo\metre} between the participants. We demonstrated the first entanglement-based QKD network achieving synchronization between the users solely by clock recovery from the photon arrival times themselves. Our system therefore neither requires special signals nor hardware for the phase alignment nor particularly stable clocks. 

The quantum bit error rate~(QBER) was automatically optimized by aligning the phases of the receiver interferometers via temperature adjustments. The receiver modules are therefore technically simple. 
Our precise method enabled building all five interferometers in a single attempt which makes fast and reliable manufacturing of higher numbers of receiver modules feasible.

We obtained sifted key rates of~\SI{29}{\bit\per\second} and QBER of~\SI{2.63}{\percent} over a total fiber length of~\SI{60.5}{\kilo\metre} as well as~\SI{6.3}{\bit\per\second} with high stability. We also obtained a QBER of~\SI{4.5}{\percent} over a total fiber length of~\SI{108}{\kilo\metre}, of which~\SI{26.8}{\kilo \meter} were field deployed fiber.

Time-~and wavelength-division multiplexing schemes demonstrated with polarization-based entanglement protocols can be applied to our scheme as well. However, in contrast to networks using polarization encoding, our setup is unaffected by environmental disturbances deteriorating the polarization in the transmission fiber. This allows for a considerable improvement in terms of the stability of secure key rates due to the reduced QBER.
These advantages of time-bin coding can be readily combined with wavelength division multiplexing for robust metropolitan-scale networks. We demonstrated demultiplexing with a regular arrayed-waveguide grating with~\SI{100}{\giga \hertz}~channel spacing as well as with a wavelength-selective switch, which enables to dynamically change the user combinations and bandwidth for user pairs. 

Currently our system is readily scalable to simultaneously provide at least 34~participants in 17~pairs with keys. With suitable~\SI{50}{\giga\hertz} demultiplexing, the size of our network can be extended to more than 100~participants grouped in pairs solely based on wavelength demultiplexing. Even higher numbers of users or networks with subnets will become feasible when additional time multiplexing is used.

\section*{Acknowledgement}

This research has been funded by the Deutsche Forschungsgemeinschaft (DFG, German Research Foundation) – SFB 1119 – 236615297

We thank Paul Wagner from Deutsche Telekom Technik GmbH for lending us the AWG and fiber spools.

\section*{Data Availability Statement}
The data that support the findings of this study are available from the corresponding author upon reasonable request.


%

\end{document}